\documentclass[aps,prl,twocolumn,showpacs,superscriptaddress,groupedaddress]{revtex4}

\usepackage{amsthm}
\usepackage{graphicx}
\usepackage{amsfonts}
\usepackage{amssymb}
\usepackage{times,txfonts}
\usepackage{hyperref}
\usepackage{color}

\begin{document}

\title{Frozen condition of quantum coherence for atoms on a stationary trajectory}

\author{Anwei Zhang}
\email{awzhang@sjtu.edu.cn}
\affiliation{ School of Physics and Astronomy, Shanghai Jiao Tong University, Shanghai 200240, China}

\begin{abstract}
We consider two co-moving atoms on a stationary trajectory and develop a formalism to characterize the properties of such atoms. We give a criterion under which quantum coherence (QC) is frozen to a nonzero value and show that the frozen condition (FC) is not so sensitive to the initial condition of state. We introduce the concept super- and subradiant spontaneous excitation rates which plays an equivalent role as conventional collective emission rates in the evolution of quantum coherence. We also give the general relationship between the quantities characterizing properties of atoms in thermal bath and show that the enhanced quantum coherence and subradiant state can be gained from initial state.
\end{abstract}


\maketitle




 \noindent\emph{\bfseries Introduction.} QC as a consequence of quantum state superposition principle is one of the key features that result in nonclassical phenomena. It is a powerful resource in quantum information theory as well as entanglement and discord-type quantum correlation. Though its importance in fundamental physics, only recently, have relevant steps been attempted to develop a rigorous framework to quantify QC for general states, such as the relative entropy of QC and the $l_1$ norm \cite{norm}.

A long-standing and significant issue concerning QC is the decoherence induced by the inevitable interaction between system and environment. In the past few years, several proposals have been suggested for fighting against the deterioration of QC, for instance, decoherence-free subspaces \cite{free2,free3}, quantum error correction codes \cite{qecc3}, dynamical decoupling \cite{dd2} or quantum Zeno dynamics \cite{zeno1,zeno2}.
Recently, the conditions of sustaining long-lived QC were investigated \cite{fro}.
It was shown that QC can remain unchanged with time (freezing coherence) and the FC for two qubits, undergoing local identical bit-flip channels with Bell-diagonal initial states, was only dependent on the initial condition of the states \cite{Streltsov}.

In realistic physical system, atoms usually cannot be handled simply as noninteracting individual qubits, when atomic spacing is small. Besides, for an ensemble of atoms, motion and temperature are also important factors which should be taken into account. Then searching for a general FC for interacting atoms under normal conditions is necessary in practice.

In this letter, we investigate the
FC of QC for two identical two-level atoms on stationary trajectory
which has a characterization that the geodesic distance between two points on trajectory depends only on the proper time interval \cite{Letaw,aud}. Thus for stationary trajectory, field correlation function is invariance under translations in time.  Besides, the stationary trajectory guarantees that the atoms have stationary states. Inertial atom in Minkowski vacuum or thermal bath and uniformly or circularly accelerated atom viewed by instantaneous inertial observer are all stationary. We find that the QC for interacting atoms on stationary trajectory can also be long lived, however the FC is not so sensitive to the initial condition of single excitation state but to super- or sub-radiant decay rate of atoms. We develop a formalism to describe atoms on stationary trajectory, and give the general relationship between the quantities characterizing properties of atoms. Besides, we show that enhanced QC and sub-radiant state can be obtained from initial state.



\begin{figure}[htbp]
\centering
\includegraphics[width=0.357\textwidth ]{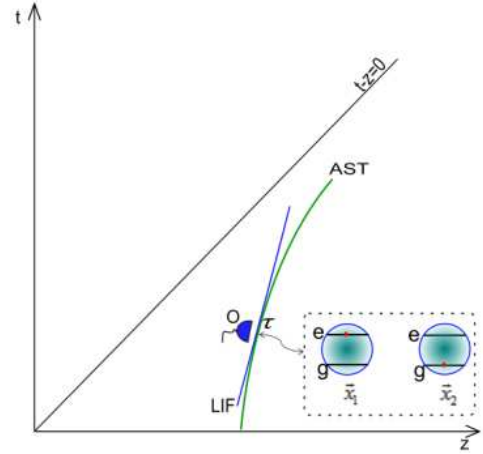}
\caption{Schematic illustration for two identical two-level atoms comoving on an arbitrary stationary trajectory (AST). The
observer (O) located in the local inertial frame (LIF) of AST  is instantaneous static with respect to the atoms.
The coordinate of atoms $(t,\vec{x}_j)$ is characterized by proper time $\tau$ as $x_j(\tau)=(t(\tau),\vec{x}_j(\tau))$.
The observer can also be in
laboratory reference frame and the difference is that the total Hamiltonian we have used should be multiplied by $d\tau/dt$.
}\label{figure.1}
\end{figure}

\smallskip
\noindent\emph{\bfseries Formalism.} We consider two identical two-level atoms moving on stationary trajectories $x_j(\tau)=(t(\tau),\vec{x}_j(\tau))$ in a fluctuating vacuum electromagnetic field, where $(t,\vec{x}_j)$ are the Minkowski coordinates of atom $j$ referring to an inertial reference frame and $\tau$ denotes proper time of these two comoving atoms (see Fig.~\ref{figure.1}). The total Hamiltonian
of the coupled system can be described by $H=H_s+H_f+H_I$.
Here $H_s$ is the free
Hamiltonian of atoms and its explicit expression in Schr\"{o}dinger picture is
$H_s=\sum^{2}_{j=1}\omega_0\sigma^{+}_j\sigma^{-}_j$,
where $\omega_0$ is the level spacing of the two-level atoms,
$\sigma^{+}_j=|e_j\rangle\langle g_j|$ and $\sigma^{-}_j=|g_j\rangle\langle e_j|$ are, respectively,
the raising and lowering operators of the atom $j$.
The free Hamiltonian $H_f$ with respect to $\tau$  takes the form
$ H_f=\sum_{k\lambda}\omega_{\vec{k}\lambda} a^{\dag}_{\vec{k}\lambda}a_{\vec{k}\lambda}\frac{dt}{d\tau}$ \cite{jj}.
Here $a^{\dag}_{\vec{k}\lambda}$, $a_{\vec{k}\lambda}$ are the creation and annihilation operators for a photon with momentum $\vec{k}$,
frequency $\omega_{\vec{k}}$ and polarization $\lambda$.
 The Hamiltonian $H_I$ that describes the atom-field interaction can be written in electric dipole approximation in $\tau$ as $-e\sum^{2}_{j=1}\vec{r}_j\cdot\vec{E}(\vec{x}_j(\tau))=-e\sum^{2}_{j=1}
 (\vec{d}_j\sigma^{+}_j+\vec{d}^{\ast}_j\sigma^{-}_j)\cdot
 \vec{E}(\vec{x}_j(\tau)),$
 where $e$ is the electron electric charge, $e\vec{r}_j$ is the electric dipole moment for atom $j$, $\vec{d}_j=\langle e_j|\vec{r}_j|g_j\rangle$ and $\vec{E}(\vec{x}_j(\tau))$ is the electric field strength evaluated at the position $\vec{x}_j(\tau)$.
The Hamiltonian in Schr\"{o}dinger picture can be changed to
interaction picture via unitary transformation with unitary operator
 $U_0(\tau)=\mathrm{exp}\big[-i\sum^{2}_{j=1}\omega_0\sigma^{+}_j\sigma^{-}_j\tau-i\sum_{k\lambda}\omega_{\vec{k}\lambda} a^{\dag}_{\vec{k}\lambda}a_{\vec{k}\lambda}t(\tau)\big]$, which is the solution to the Schr\"{o}dinger equation in $\tau$:
$i\frac{d}{d\tau}U_0(\tau)=(H_s+H_f)U_0(\tau).$
Then the atom-field interaction Hamiltonian in the interaction picture can be written as
\begin{equation}\label{hami}
H_I(\tau)=-e\sum^{2}_{j=1}
 (\vec{d}\sigma^{+}_j e^{i\omega_0\tau}+\vec{d}^{\ast}\sigma^{-}_je^{-i\omega_0\tau})\cdot\vec{E}(x_j(\tau)),
\end{equation}
where $\vec{E}(x_j(\tau))=U^{\dagger}_0(\tau)\vec{E}(\vec{x}_j(\tau))U_0(\tau)$ and we have let $\vec{d}_1=\vec{d}_2=\vec{d}$ for simplicity.

Here we consider the polarizations of the coupling photon required by the two atoms only in the same direction and for the convenience of calculations
 we assume (i)
\begin{equation}\label{as1}
  G^{11}_{ii}(u)=G^{22}_{ii}(u),\;G^{12}_{ii}(u)=G^{21}_{ii}(u),
\end{equation}
where $G^{ab}_{ij}(\tau-\tau^{'})=
\langle0| {E}^{+}_i(x_a(\tau)){E}^{-}_j(x_b(\tau^{'}))|0\rangle$ is the electric field correlation function and we have decomposed
$\vec{E}(x_j(\tau))$ in $H_I(\tau)$ into positive- and negative-frequency parts:
$\vec{E}(x_j(\tau))=\vec{E}^{+}(x_j(\tau))+\vec{E}^{-}(x_j(\tau))$
with $\vec{E}^{+}(x_j(\tau))|0\rangle=0$ and $\langle0|\vec{E}^{-}(x_j(\tau))=0$. Under such an assumption, the correlation function is invariant under the exchange of the two atoms, thus there is no difference in the atom-field interaction between the atoms and the external environment for thm is same.
We further assume (ii) the interaction between atoms and field to be weak, so the Wigner-Weisskopf approximation can be adopt.

Consider the atoms with initial single excitation state and field as vacuum state
 $|\varphi(0)\rangle=\cos\frac{\theta}{2}|e_1g_2\rangle|0\rangle+
 \sin\frac{\theta}{2}|g_1e_2\rangle|0\rangle$.
 Then at time $\tau$, the general form of the state vector can be written as \cite{Stephen,1976}
 \begin{eqnarray}\label{8}
|\varphi(\tau)\rangle&=&
\sum_{k\lambda}b_{\vec{k}\lambda1}(\tau)|g_1g_2\rangle|1_{\vec{k}\lambda}\rangle
+\sum_{k\lambda}b_{\vec{k}\lambda2}(\tau)|e_1e_2\rangle|1_{\vec{k}\lambda}\rangle \nonumber \\
&&+b_1(\tau)|e_1g_2\rangle|0\rangle+b_2(\tau)|g_1e_2\rangle|0\rangle,
 \end{eqnarray}
where  $|1_{\vec{k}\lambda}\rangle$ denotes one photon in the mode $(\vec{k},\lambda)$. It is worth noting that
this state is observed in local inertial reference frame of atoms.

The state probability amplitudes in (\ref{8}) can be obtained (see the Supplemental Material \cite{SM}):
\begin{eqnarray}\label{abb}
 b_1(\tau) &=&\frac{1}{2}\big[(\cos\frac{\theta}{2}+\sin\frac{\theta}{2})C_{+}(\tau)
 +(\cos\frac{\theta}{2}-\sin\frac{\theta}{2})C_{-}(\tau)\big],\nonumber \\ b_2(\tau)&=&\frac{1}{2}\big[(\cos\frac{\theta}{2}+\sin\frac{\theta}{2})C_{+}(\tau)
 -(\cos\frac{\theta}{2}-\sin\frac{\theta}{2})C_{-}(\tau)\big],\nonumber \\
\end{eqnarray}
where $C_{\pm}(\tau)=e^{-(A^{11}(0)+B^{11}(0)\pm A^{12}(0)\pm B^{12}(0))\tau}$.
And  $A^{ab}(0)$ and $B^{ab}(0)$ can be decomposed as follows
\begin{eqnarray}\label{av}
A^{ab}(0)&=& \frac{1}{2}e^{2}d_id^{\ast}_i\mathcal{G}^{ab}_{ii}(\omega_0)+
  ie^{2}d_id^{\ast}_i\mathcal{K}^{ab}_{ii}(\omega_0),\nonumber \\
B^{ab}(0)&=&\frac{1}{2}e^{2}d^{\ast}_id_i\mathcal{G}^{ab}_{ii}(-\omega_0)+
ie^{2}d^{\ast}_id_i\mathcal{K}^{ab}_{ii}(-\omega_0)
 \end{eqnarray}
with
\begin{eqnarray}\label{b1}
\mathcal{G}^{ab}_{ii}(\pm\omega_0)&=&\int^{\infty}_{-\infty}due^{\pm i\omega_0 u}
G^{ab}_{ii}(u),\nonumber \\
\mathcal{K}^{ab}_{ii}(\pm\omega_0)&=&-\frac{P}{2\pi }\int^{\infty}_{-\infty}\frac{\mathcal{G}^{ab}_{ii}(\omega)}{\omega\mp\omega_0}d\omega.
\end{eqnarray}
Here $P$ denotes the Cauchy principal value. It is worthwhile to note that $e^{2}d_id^{\ast}_i\mathcal{G}^{11}_{ii}(\omega_0)$, $e^{2}d^{\ast}_id_i\mathcal{G}^{11}_{ii}(-\omega_0)$ are the spontaneous
emission rate $\Gamma^{11}_\downarrow$ and spontaneous excitation rate $\Gamma^{11}_\uparrow$ of atom $1$, respectively \cite{anwei}.
$e^{2}d_id^{\ast}_i\mathcal{G}^{12}_{ii}(\omega_0)$ and $e^{2}d^{\ast}_id_i\mathcal{G}^{12}_{ii}(-\omega_0)$ can be regarded as the
modulation of the spontaneous emission rate $\Gamma^{12}_\downarrow$  \cite{dip} and spontaneous excitation rate $\Gamma^{12}_\uparrow$  of one atom due to the presence of another atom. And $\Gamma^{11}_\downarrow\pm\Gamma^{12}_\downarrow$ are actually the super- and sub-radiant  spontaneous emission rate $\Gamma_{\downarrow\pm}$, respectively \cite{aw}.
$\Gamma^{11}_\uparrow\pm\Gamma^{12}_\uparrow$ can be respectively termed as super- and sub-radiant spontaneous excitation rate $\Gamma_{\uparrow\pm}$. The clear definitions and the relations between above quantities are listed in Table ~\ref{I}.
$e^{2}d_id^{\ast}_i\mathcal{K}^{aa}_{ii}(\omega_0)$ and $e^{2}d^{\ast}_id_i\mathcal{K}^{aa}_{ii}(-\omega_0)$ represent the level shift of upper state and lower state of atom $a$. Finally, $e^{2}d_id^{\ast}_i\mathcal{K}^{12}_{ii}(\omega_0)+e^{2}d^{\ast}_id_i\mathcal{K}^{12}_{ii}(-\omega_0)$ is the dipole-dipole interaction potential $V$, which results from photon exchanges between atoms.

\noindent\emph{\bfseries Frozen condition.} Now we apply the previously developed formalism to investigate QC for two atoms on stationary trajectory.
For simplicity, we take $l_1$ norm \cite{norm} as a measure of QC, which is defined as $\mathcal{C}_{l_1}(\rho)=\sum_{i\neq j}|\rho_{i,j}|.$
The reduced density matrix $\rho$ obtained by tracing the density matrix of the total system over the field degrees of freedom can be written in the basis of the product states, $|1\rangle=|e_1e_2\rangle$, $|2\rangle=|e_1g_2\rangle$, $|3\rangle=|g_1e_2\rangle$, $|4\rangle=|g_1g_2\rangle$. Since the state $|e_1e_2\rangle|1_{\vec{k}\lambda}\rangle$ is only an intermediate state and short
lived, the off-diagonal elements $\rho_{14}$ and $\rho_{41}$ take effect only in short time. However we are interested in the long time behavior of QC, so these cross terms can be neglected. Then QC will be simply expressed as
$\mathcal{C}_{l_1}(\rho)=2|b_1(\tau)b^{\ast}_2(\tau)|
=e^{-\Gamma^{11}\tau}
\sqrt{\cos^{2}\theta\sin^{2}(2V\tau)+[\sin\theta\cosh(\Gamma^{12}\tau)
-\sinh(\Gamma^{12}\tau)]^{2}},$ which depends only on $\rho_{23}$.
Here $\Gamma^{11}=\Gamma^{11}_\downarrow+\Gamma^{11}_\uparrow$ and $\Gamma^{12}=\Gamma^{12}_\downarrow+\Gamma^{12}_\uparrow$, see Table~\ref{I}.
\begin{table}[htbp]
\begin{tabular}{|p{53 pt}| p{62 pt}| p{68 pt}|p{41 pt}|}
 \hline
 &Emission & Excitation & Plus \\
 \hline
Spontaneous  transition rates & $\Gamma^{11}_\downarrow=D^{2}_{ii}\mathcal{G}^{11}_{ii}(\omega_0)$ & $\Gamma^{11}_\uparrow=D^{2}_{ii}\mathcal{G}^{11}_{ii}(-\omega_0)$ &$\Gamma^{11}_\downarrow+\Gamma^{11}_\uparrow=\Gamma^{11}$\\
\hline
Corresponding modulations & $\Gamma^{12}_\downarrow=D^{2}_{ii}\mathcal{G}^{12}_{ii}(\omega_0)$ & $\Gamma^{12}_\uparrow=D^{2}_{ii}\mathcal{G}^{12}_{ii}(-\omega_0)$ & $\Gamma^{12}_\downarrow+\Gamma^{12}_\uparrow=\Gamma^{12}$\\
\hline
Super-/sub-radiant rates (Plus/Minus) & $\Gamma_{\downarrow\pm}=\Gamma^{11}_\downarrow \pm \Gamma^{12}_\downarrow$ & $\Gamma_{\uparrow\pm}=\Gamma^{11}_\uparrow\pm\Gamma^{12}_\uparrow$ & $\Gamma^{11}\pm\Gamma^{12}=\Gamma_{\downarrow\pm}+\Gamma_{\uparrow\pm}$\\
\hline
\end{tabular}
 \centering\caption{The Relationship between the quantities characterizing atomic properties.
  Here Plus/Minus denotes the sum or difference of the previous two terms and $D^{2}_{ii}=e^{2}d_id^{\ast}_i$.}\label{I}
\end{table}

For the initial sub-radiant state $(|e_1g_2\rangle-|g_1e_2\rangle)|0\rangle/\sqrt 2$, QC will decay exponentially as $e^{-(\Gamma^{11}-\Gamma^{12})\tau}$. Then it can be frozen to maximum value $1$ only in the case that $\Gamma^{11}-\Gamma^{12}=0$.
When the initial state is a separable state, that is $|e_1g_2\rangle|0\rangle$ or $|g_1e_2\rangle|0\rangle$, QC will increase from zero and evolve as $e^{-\Gamma^{11}\tau}\sqrt{\sin^{2}(2V\tau)+\sinh^{2}(\Gamma^{12}\tau)}$. After evolving for a sufficiently long time $\tau\gg 1/\Gamma^{11}$, it is frozen to $1/2$ if $\Gamma^{11}-\Gamma^{12}=0$.
Generally, one can find that when $\tau$ is much larger than $1/\Gamma^{11}$, QC will evolve to
a nonzero constant $(1-\sin\theta)/2$ ($\theta\neq \pi/2$) under the condition that $\Gamma^{11}-\Gamma^{12}=0$. Such a
FC can be rewritten as $\Gamma_{\downarrow-}+\Gamma_{\uparrow-}=0$, which means that the sum of sub-radiant spontaneous emission and excitation rate (namely sub-radiant decay rate) should be zero. For inertial atoms in Minkowski vacuum, there is no spontaneous excitation ($\Gamma^{11}_\uparrow$ and $\Gamma^{12}_\uparrow$ all vanish), the FC will be simplified to $\Gamma_{\downarrow-}=0$ that the sub-radiant spontaneous emission rate is null.
In the above, we only consider the case that sub-radiant decay rate equals zero under which one can freeze QC except for the initial state with $\theta=\pi/2$.
Because this state is super-radiant state, and QC decays as $e^{-(\Gamma^{11}+\Gamma^{12})\tau}$ which can be frozen only
when super-radiant decay rate vanishes.
In physical implementations, when super-radiant decay rate vanishes, the sub-radiant decay rate will usually vanish too.
In such a case, the QC for arbitrary initial state will all be frozen.

\smallskip
\noindent\emph{\bfseries Thermal bath.} How would the above results have been modified if the initial state of field were not vacuum, but a thermal bath described by density matrix $\rho=e^{-\beta H_{f}}$ with $\beta$ being the inverse temperature? Since the thermal equilibrium state is a stationary state, the above formalism can be easily generalize to this case. Following the similar procedure, it can be found that, for atoms at rest in thermal bath, the correlation function in (\ref{b1}) should be replaced by
thermal Green function $G^{ab}_{ii\beta}(t-t^{'})$ which is expressed as $\langle  {E}^{+}_i(x_a(t)){E}^{-}_i(x_b(t^{'}))\rangle_\beta+\langle  {E}^{-}_i(x_a(t)){E}^{+}_i(x_b(t^{'}))\rangle_\beta=\mathrm{Tr}\big(\rho {E}_i(x_a(t)){E}_i(x_b(t^{'}))\big)$. Then $\Gamma^{11}$ and $\Gamma^{12}$ in initial vacuum state case are replaced by $\Gamma^{11}_\beta=\Gamma^{11}_{\downarrow\beta}+\Gamma^{11}_{\uparrow\beta}$ and $\Gamma^{12}_\beta=\Gamma^{12}_{\downarrow\beta}+\Gamma^{12}_{\uparrow\beta}$, where $\Gamma^{11}_{\downarrow\beta}=e^{2}d_id^{\ast}_i\mathcal{G}^{11}_{ii\beta}(\omega_0)$ has the meaning of total emission rate including spontaneous
emission and stimulated radiation rate, $\Gamma^{11}_{\uparrow\beta}=e^{2}d^{\ast}_id_i\mathcal{G}^{11}_{ii\beta}(-\omega_0)$ has the meaning of absorption rate \cite{anwei} and $\Gamma^{12}_{\downarrow\beta}=e^{2}d_id^{\ast}_i\mathcal{G}^{12}_{ii\beta}(\omega_0)$, $\Gamma^{12}_{\uparrow\beta}=e^{2}d^{\ast}_id_i\mathcal{G}^{12}_{ii\beta}(-\omega_0)$ can be regarded as the corresponding modulations.

Now the FC of QC has to be changed to $\Gamma^{11}_\beta-\Gamma^{12}_\beta$ (or $\Gamma^{11}_\beta+\Gamma^{12}_\beta$)=$0$. Next we will simplify this condition.
For thermal Green function, it can be verified that
$ G^{ab}_{ii\beta}(t-t^{'})=G^{ba}_{ii\beta}(t^{'}-t-i\beta).$
Actually,  this is Kubo-Martin-Schwinger (KMS) condition and thermal state is a KMS state.
Taking Fourier transforms of KMS condition gives
\begin{equation}\label{32}
  \Gamma^{ab}_{\downarrow\beta}=e^{\omega_0\beta}\Gamma^{ab}_{\uparrow\beta}.
\end{equation}
Here we have utilized $G^{ab}_{ii\beta}(u)=G^{ba}_{ii\beta}(u)$ which is equivalent to the assumption (i).
Due to the fact that the commutator of field is a c-number, its expectation values should be independent of the field state:
\begin{equation}\label{33}
  \langle  [{E}_i(x_a(t)),{E}_i(x_b(t^{'}))]\rangle_\beta= \langle 0| [{E}_i(x_a(t)),{E}_i(x_b(t^{'}))]|0\rangle.
\end{equation}
The Fourier transforms of this equation leads to
\begin{equation}\label{34}
  \Gamma^{ab}_{\downarrow\beta}-\Gamma^{ab}_{\uparrow\beta}= \Gamma^{ab}_{\downarrow}.
\end{equation}
Note that we have omitted the term $\Gamma^{ab}_{\uparrow}$, since there is no spontaneous excitation for static atoms in Minkowski thermal bath. (\ref{32}) and (\ref{34}) give
\begin{equation}\label{35}
  \Gamma^{ab}_{\downarrow\beta}=\frac{e^{\omega_0\beta}}{e^{\omega_0\beta}-1}\Gamma^{ab}_{\downarrow},\;
  \Gamma^{ab}_{\uparrow\beta}=\frac{1}{e^{\omega_0\beta}-1}\Gamma^{ab}_{\downarrow}.
\end{equation}
Thus the FC can be simplified to $\Gamma_{\downarrow-}$ (or $\Gamma_{\downarrow+}$)=$0$.
This FC is irrelevant to temperature, thus the presence of thermal bath or not does not change
the FC of QC for static atoms. However if FC is not satisfied, QC will decay faster as temperature increases, since the decay rate is enhanced $(1+2n)$ times compared with zero temperature case (see Table~\ref{II}).
\begin{table}[htbp]
\begin{tabular}{|p{60 pt}| p{60 pt}| p{45 pt}|p{60 pt}|}
 \hline
 &Total emission & Absorption & Plus \\
 \hline
 Transition rates of atom 1 & $\Gamma^{11}_{\downarrow\beta}=(1+n)\Gamma^{11}_\downarrow$ & $\Gamma^{11}_{\uparrow\beta}=n\Gamma^{11}_\downarrow$ &$\Gamma^{11}_\beta=(1+2n)\Gamma^{11}_\downarrow$\\
\hline
Corresponding modulations & $\Gamma^{12}_{\downarrow\beta}=(1+n)\Gamma^{12}_\downarrow$ & $\Gamma^{12}_{\uparrow\beta}=n\Gamma^{12}_\downarrow$ & $\Gamma^{12}_\beta=(1+2n)\Gamma^{12}_\downarrow$\\
\hline
Plus/Minus & $(1+n)\Gamma_{\downarrow\pm}$ & $n\Gamma_{\downarrow\pm}$ & $\Gamma^{11}_\beta\pm\Gamma^{12}_\beta=(1+2n)\Gamma_{\downarrow\pm}$\\
\hline
\end{tabular}
\centering\caption{Quantities characterizing the properties of atoms in thermal bath, where $n=1/(e^{\omega_0\beta}-1)$.}\label{II}
\end{table}

\noindent\emph{\bfseries Discussion.} Due to the fact that the Hamiltonian used above is invariant in form under the presence of boundaries in space, thus our results are applicable to not only atoms in free space but also that in bounded space, as long as the assumptions (i) and (ii) are satisfied.

Why can the QC be frozen at a nonzero value? For the state (\ref{8}), after long time evolution, it is usually a single-photon state, then the
measure of QC will be finally zero. But in extreme condition, such as, only sub-radiant decay rate vanishes, the state will be a
superposition of $|e_1g_2\rangle|0\rangle$ with probability $(1-\sin\theta)/4$, $|g_1e_2\rangle|0\rangle$ with the same probability and single-photon state. The QC is thus preserved. Actually in such a case, the system as a whole does not decay any more and evolves into a steady state. So QC can be frozen.

 When the atomic separation is much less than the resonant radiation wavelength of atoms, that is the Dicke limit \cite{dicke}, the sub-radiant spontaneous emission and excitation rates will all approach to zero, as can be seen from their definitions. The FC is thus satisfied.
For implementation, the long-wavelength atoms or molecules,
such as Rydberg atoms \cite{ryd}, are appropriate choices.

Another optional strategy is to place the atoms very near the surface of plates. Due to the fact that the tangential component of fluctuating vacuum electric field on the boundary is null, so when the distance from atoms to boundary is much less than resonant radiation wavelength of atoms and the polarization direction of atoms is in the surface, electric field correlation function will go to zero. Thus in this case, super- and sub-radiant decay rates will all tend to zero, FC is satisfied.

 \begin{figure}[htbp]
\centering
\includegraphics[width=0.237\textwidth ]{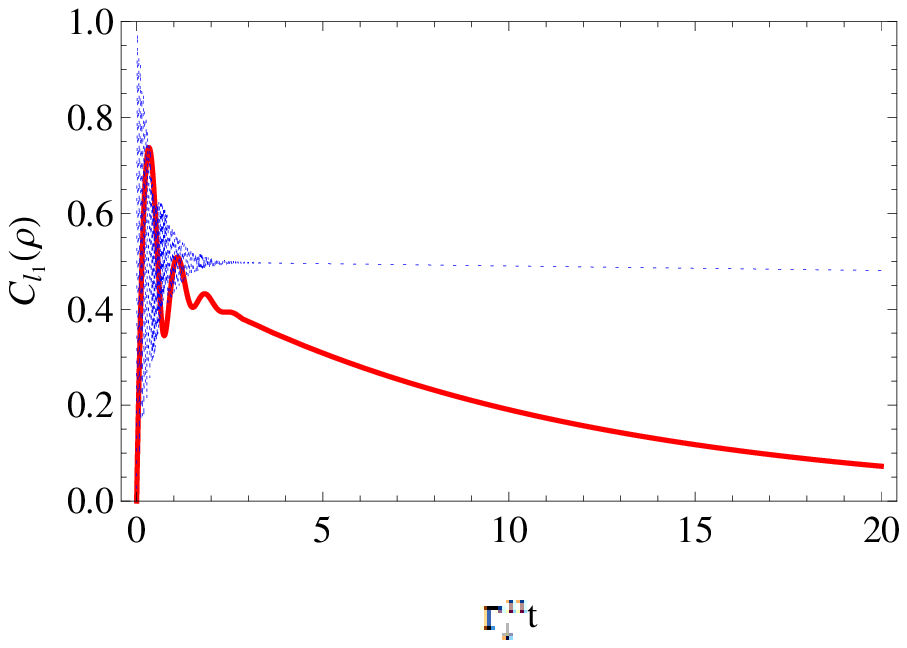}
\includegraphics[width=0.237\textwidth ]{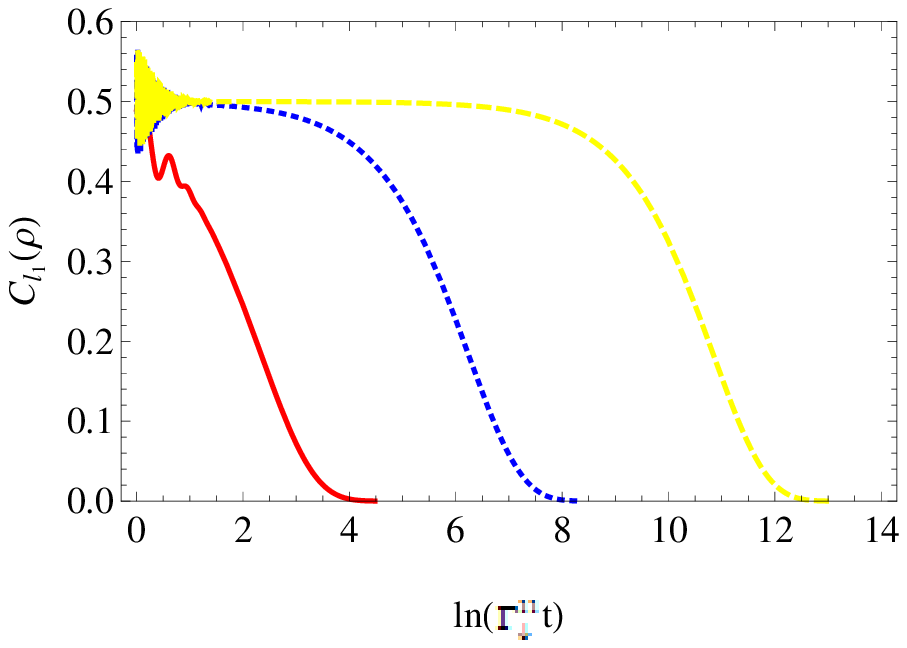}
\caption{ The measure of coherence for static atoms with initial separable state in free space as a function of $\Gamma^{11}_\downarrow t$ and $\ln(\Gamma^{11}_\downarrow t)$, respectively.
In such a case, the modulation $\Gamma^{12}_\downarrow=3\Gamma^{11}_\downarrow\frac{\sin R-R \cos R}{R^{3}}$ and the interaction potential
$V=-3\Gamma^{11}_\downarrow\frac{\cos R+R \sin R}{2R^{3}}$
 with $R=r \omega_0/c$ and $r$ being atomic separation.
The blue dotted line is the $R=0.14$ case. For comparison, the red solid line with $R=1$ and yellow dashed line with $R=0.014$ are plotted.}
\label{Figure.2}
\end{figure}
Next we take static atoms in free space with polarizations along their separation as an example and plot the Fig.~\ref{Figure.2}.
It is shown that when interatomic distance is small enough compared with resonant radiation wavelength, QC is approximately frozen to 1/2. And
the shorter the atomic separation $r$, the longer the QC lives.
If we use LiH with $\omega_{0}=4.21\times10^{13}$Hz \cite{buh}, the frozen case illustrated (blue dotted line) can be fulfilled by taking $r=1\mu m$. For RbCs with $\omega_{0}=1.48\times10^{12}$Hz \cite{buh}, $r$ is large to $28\mu m$.  If we instead use Rydberg atoms, the required separation will be largely increased, since the wave emitted by Rydberg atoms can be radio-frequency or micro wave ($3\times10^{8}-3\times10^{11}$Hz).  Then by shortening the distance between such atoms, QC will be more precisely frozen and maintain for a longer time, see the case of $R=0.014$.

Now we consider the situation that these two atoms are with an acceleration $a$ perpendicular to their separation. When $\frac{a}{\omega_0c }\ll1$, in its first-order approximation, super- and sub-radiant decay rate are all enhanced $1+\frac{a}{\pi\omega_0c}$ times, and $V$ is approximately unchanged. So the FC is the same as the static case. Low acceleration does not affect FC, as the function of thermal bath.
If QC is not totally frozen, coherence will deteriorate more severely as acceleration increases. Taking initial separable state as an example, when $R=0.14$, $\Gamma^{11}_{0\downarrow} \tau=10$ ($\Gamma^{11}_{0\downarrow}$ is
spontaneous emission rate of static atom), and $\frac{a}{\pi\omega_0c}=0.01$, $0.05$ respectively, the measure of coherence will be correspondingly $0.4902$ and $0.4898$, which are all less than $0.4903$ at static case.

One may wonder why the QC for atoms initial prepared in separable state can be created to a nonzero value. Actually this can be attributed to the interaction (\ref{hami}) between atoms and environment, which makes the transitions $|e_1g_2\rangle|0\rangle \rightleftarrows |e_1e_2\rangle|1_{\vec{k}\lambda}\rangle \rightleftarrows |g_1e_2\rangle|0\rangle \rightleftarrows |g_1g_2\rangle|1_{\vec{k}\lambda}\rangle \rightleftarrows |e_1g_2\rangle|0\rangle$ possible.
For the initial separable state $|e_1g_2\rangle|0\rangle$,
at the neighborhood of initial time, the probability of appearing
the state $|g_1e_2\rangle|0\rangle$, $p_{1}=e^{-\Gamma^{11}_\downarrow t}[\cosh(\Gamma^{12}_\downarrow t)-\cos(2V t)]/2$, increases from zero. The interference term $\rho_{23}$, with $|\rho_{23}|=\sqrt{p_{1}p_{2}}$ and $p_{2}=e^{-\Gamma^{11}_\downarrow t}[\cosh(\Gamma^{12}_\downarrow t)+\cos(2V t)]/2$ being the probability of appearing the state $|e_1g_2\rangle|0\rangle$,
comes up. Since then the system evolves into a superposition state, and the QC varies with the change of the probability distribution of each state.

For a general initial state, the measure of coherence is $|\sin\theta|$. It will be less than the frozen value $(1-\sin\theta)/2$ $(\theta\neq\pi/2)$ in the case that $\sin\theta \in (-1, 1/3)$. Thus for a initial state in this range, the QC can be enhanced by engineering the sub-radiant decay rate. When sub-radiant decay rate is small enough and $\tau\gg 1/(\Gamma^{11}+\Gamma^{12})$, the state will be a sub-radiant state, which can be easily found in the coupled basis $\{|e_1e_2\rangle, |S\rangle=(|e_1g_2\rangle+|g_1e_2\rangle)/\sqrt{2}, |A\rangle=(|e_1g_2\rangle-|g_1e_2\rangle)/\sqrt{2}, |g_1g_2\rangle\}$. Thus we can prepare sub-radiant state from initial state except $|S\rangle$ which is orthogonal to $|A\rangle$.

Note that the QC enhancement results from the adjustment of probability distribution induced by interaction rather than the transformation
of basis state space. Besides the QC is not inflated, but underestimated especially at the beginning of states evolution. Because the QC depends on not only the interference term $\rho_{23}$, but also $\rho_{14}$ which is omitted due to its short time behavior. Then if we initially prepare a superposition state of $|e_1g_2\rangle$ and $|g_1e_2\rangle$,
after long time evolution of the states, the QC will still be only encoded on these two states.

What is the relation between QC being investigated and entanglement \cite{mehmet}?
For our X form density matrix, the concurrence \cite{Wootters} as a measure of entanglement, is Max$\{0,2(|\rho_{23}|-\sqrt{\rho_{11}\rho_{44}}),2(|\rho_{14}|-\sqrt{\rho_{22}\rho_{33}})\}$ \cite{Ikram}.
We can see that the concurrence is not greater than QC, $2(|\rho_{23}|+|\rho_{14}|)$. But when time is large enough, $\rho_{14}$ and $\rho_{11}$ all approximate zero,  concurrence will tend to QC.

 Our results can be generalized to many atoms case. Our formalism can be used to investigate the decoherence induced by gravity \cite{Pikovski} or explore the structure of spacetime \cite{Steeg}.

\begin{acknowledgments}
A. Z. would like to thank K. Zhang for giving some
suggestions and polishing three to four sentences. A. Z. slao wants to thank R.W. Field for providing some
information, as well as Fangwei Ye, Xianfeng Chen and
Ying Xue for help in changing tutor. 
\end{acknowledgments}



\end{document}